\documentclass[aps,prl,twocolumn,superscriptaddress]{revtex4-1}

\usepackage[dvipdf]{graphicx}
\usepackage{dcolumn}
\usepackage{bm}
\usepackage[reqno, intlimits,]{amsmath}
\usepackage[latin1]{inputenc}
\usepackage{textcomp}

\begin{document}

\title{Evidence of resonant surface wave excitation in the relativistic regime through measurements of proton acceleration from grating targets}

\author{T.~Ceccotti}\email[ ]{tiberio.ceccotti@cea.fr}
\author{V.~Floquet}
\affiliation{CEA/IRAMIS/SPAM, F-91191 Gif-sur-Yvette, France}
\author{A.~Sgattoni}
\affiliation{Istituto Nazionale di Ottica, Consiglio Nazionale delle Ricerche, research unit ``Adriano Gozzini'', Pisa, Italy}
\affiliation{Dipartimento di Energia, Politecnico di Milano, Milano, Italy}
\author{A.~Bigongiari}
\affiliation{LULI, Universit\'{e} Pierre et Marie Curie, Ecole Polytechnique, CNRS, CEA, 75252 Paris, France}
\author{O.~Klimo}
\affiliation{FNSPE, Czech Technical University in Prague, CR-11519 Prague, Czech Republic}
\affiliation{Institute of Physics of the ASCR, ELI-Beamlines project, Na Slovance 2, 18221 Prague, Czech Republic}
\author{M.~Raynaud}
\affiliation{CEA/DSM/LSI, CNRS, Ecole Polytechnique, 91128 Palaiseau Cedex, France}
\author{C.~Riconda}
\affiliation{LULI, Universit\'{e} Pierre et Marie Curie, Ecole Polytechnique, CNRS, CEA, 75252 Paris, France}
\author{A.~Heron}
\affiliation{CPHT, CNRS, Ecole Polytechnique, 91128 Palaiseau Cedex, France}
\author{F.~Baffigi}
\author{L.~Labate} 
\author{L.~A.~Gizzi}
\affiliation{Istituto Nazionale di Ottica, Consiglio Nazionale delle Ricerche, research unit ``Adriano Gozzini'', Pisa, Italy}
\author{L.~Vassura}
\affiliation{LULI, UMR7605, CNRS-CEA-Ecole Polytechnique-Paris 6, 91128 Palaiseau, France}
\affiliation{Dipartimento SBAI, Universit\`a di Roma ``La Sapienza'', Via A. Scarpa 14, 00161 Roma, Italy}
\author{J.~Fuchs}
\affiliation{LULI, UMR7605, CNRS-CEA-Ecole Polytechnique-Paris 6, 91128 Palaiseau, France}
\author{M.~Passoni}
\affiliation{Dipartimento di Energia, Politecnico di Milano, Milano, Italy}
\author{M.~Kv\v{e}ton}
\author{F.~Novotny}
\author{M.~Possolt}
\affiliation{FNSPE, Czech Technical University in Prague, CR-11519 Prague, Czech Republic}
\author{J.~Prok\r{u}pek}
\affiliation{FNSPE, Czech Technical University in Prague, CR-11519 Prague, Czech Republic}
\affiliation{Institute of Physics of the ASCR, ELI-Beamlines project, Na Slovance 2, 18221 Prague, Czech Republic}
\author{J.~Pro\v{s}ka}
\affiliation{FNSPE, Czech Technical University in Prague, CR-11519 Prague, Czech Republic}
\author{J.~P\v{s}ikal}
\author{L.~\v{S}tolcov\'{a}}
\affiliation{FNSPE, Czech Technical University in Prague, CR-11519 Prague, Czech Republic}
\affiliation{Institute of Physics of the ASCR, ELI-Beamlines project, Na Slovance 2, 18221 Prague, Czech Republic}
\author{A.~Velyhan}
\affiliation{Institute of Physics of the ASCR, ELI-Beamlines project, Na Slovance 2, 18221 Prague, Czech Republic}
\author{M.~Bougeard}
\author{P.~D'Oliveira}
\author{O.~Tcherbakoff}
\author{F.~R\'{e}au}
\author{P.~Martin}
\affiliation{CEA/IRAMIS/SPAM, F-91191 Gif-sur-Yvette, France}
\author{A.~Macchi}\email[ ]{andrea.macchi@ino.it}
\affiliation{Istituto Nazionale di Ottica, Consiglio Nazionale delle Ricerche, research unit ``Adriano Gozzini'', Pisa, Italy}
\affiliation{Dipartimento di Fisica ``Enrico Fermi'', Universit\`a di Pisa, Largo Bruno Pontecorvo 3, I-56127 Pisa, Italy}

\date{\today}

\begin{abstract}
The interaction of laser pulses with thin grating targets, having a periodic groove at the irradiated surface, has been experimentally investigated. Ultrahigh contrast ($\sim 10^{12}$) pulses allowed to demonstrate an enhanced laser-target coupling for the first time in the relativistic regime of ultra-high intensity $>10^{19}~\mbox{W/cm}^{2}$. A maximum increase by a factor of $2.5$ of the cut-off energy of protons produced by Target Normal Sheath Acceleration has been observed with respect to plane targets, around the incidence angle expected for resonant excitation of surface waves.   
A significant enhancement is also observed for small angles of incidence, out of resonance.
\end{abstract}


\maketitle

An efficient coupling between high-intensity laser pulses and solid targets with sharp density profiles is the key to several applications, such as ion acceleration via the Target Normal Sheath Acceleration mechanism (TNSA) \cite{borghesiFST06,*daidoRPP12,*macchiRMP13}, production of coherent and incoherent X-rays \cite{tothPoP07,*chenPRL08,*levyAPL10,*zhangOE11}, isochoric heating and creation of warm dense matter \cite{doboszPRL05,*osterholzPRL06,*perezPRL10}, and studies of electron transport \cite{santosPRL02,*kosterPPCF09,*mckennaPRL11}. 
The laser-plasma interaction at the target surface is strongly sensitive to both the longitudinal profile and transverse modulations of the density on the scale of the laser wavelength (or even a smaller scale). 
Femtosecond laser pulses may be short enough that the surface structuring is preserved during the interaction and not washed away by hydrodynamical expansion, 
allowing a more efficient coupling and enhancing particle and radiation emission \cite{kulcsarPRL00,*bagchiAPB07,*sumerukPRL07,*kneipHEDP08,*chakravartyAPB11,*mondalPRB11,*ziglerPRL11}. 
In particular, targets with a periodic surface modulation (gratings) allow the resonant coupling of the laser pulse with surface waves (SWs) \cite{raether-book}
as it is widely used in plasmonics applications at low laser intensity \cite{maierJAP05}. 
So far, however, most of the studies on structured targets and on SW-induced absorption \cite{gauthierSPIE95,*kahalyPRL08,*huPoP10,*bagchiPoP12} have been limited to intensities $I \lesssim 10^{16}~\mbox{W/cm}^{2}$ because of the effect of ``prepulses'', typical of chirped pulse amplification (CPA) laser systems, which lead to early plasma formation and destruction of surface structures before the main pulse. 
Techniques such as the plasma mirror \cite{dromeyRSI04,*thauryNP07} to achieve ultrahigh pulse-to-prepulse contrast ratios now offer the opportunity to extend such studies at very high intensity.
Recently, effects of a periodic grating structure on high harmonic generation have been experimentally demonstrated at $I>10^{20}~\mbox{W/cm}^{2}$ \cite{cerchezPRL13}.

There is no detailed nonlinear theory of SWs in the regime where relativistic effects may become dominant. However, SW coupling at high intensity has been observed in particle-in-cell (PIC) simulations of laser interaction with a grating target (designed for resonant SW excitation according to linear, non-relativistic theory) which showed a strong enhancement of both absorption and energetic electron production \cite{raynaudPoP07,bigongiariPoP11,bigongiariPoP13} and, in turn, higher energies for the protons accelerated by TNSA. Thus, besides the interest of using grating targets for more efficient TNSA, the latter also provides a diagnostic for the study of SW-enhanced absorption. 

\begin{figure}[b]
\centering
\includegraphics[width=0.48\textwidth]{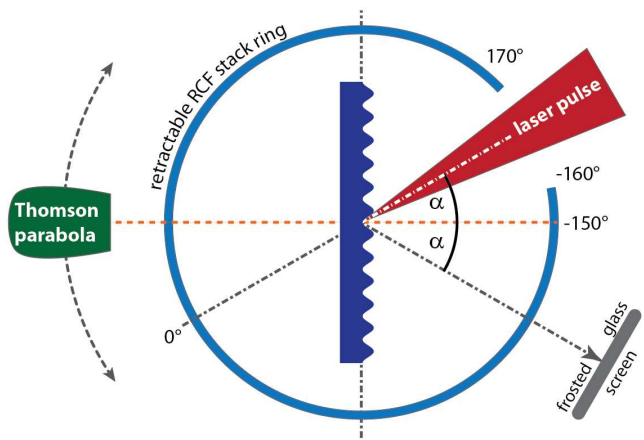}
\caption{Schematic top-view of the experimental set-up. The Thomson parabola and the frosted glass screen were located along the target normal direction and the laser specular direction respectively.
\label{setup_h}}
\end{figure}

This Letter reports on an experimental study of ultrashort laser interaction with grating targets in conditions of relativistically strong intensity ($>10^{19}~\mbox{W/cm}^{2}$) and very high contrast ($\sim10^{12}$).
The coupling enhancement was detected through simultaneous single shot measurements of TNSA proton emission and of target laser pulse reflection, as a function of the incidence angle and for different laser polarization and target thickness. 
The data show a peak of proton cut--off energies and a drop in the target reflectivity for $P$-polarized pulses when the incidence angle is close to the resonant value for SW excitation. A significant enhancement of proton energy with respect to flat targets is also observed for small incidence angles, far from resonance. The analysis is supported by 2D particle-in-cell (PIC) simulations.

The experiment was performed at the Saclay Laser Interaction Center (SLIC) facility, using the UHI100 laser delivering 80 TW ultrashort pulses (25~fs) at a central wavelength of 790 nm. 
The contrast of the beam was raised to about $10^{12}$ (high contrast, HC) thanks to a double plasma mirror \cite{levy1, kapteyn} whereas the focal spot was optimized through the correction of the laser pulse wavefront using a deformable mirror. 
The beam was focused using an off-axis $f=300~\mbox{mm}$ parabola, with 40\% of total laser energy enclosed on a spot size of $10~\mu\mbox{m}$ (diameter at $1/e^{2}$), corresponding to an intensity of about $2.5 \times 10^{19}~\mbox{W cm}^{-2}$. 

Grating targets (GTs) were produced by heat embossing into mylar foils. 
Three different values of the foil thickness (23, 40 and 0.9~$\mu$m) 
and two values of the peak-to-valley depth (0.5 and 0.3~$\mu$m) were tested.
The grating period was $d=2\lambda$ corresponding to a resonant angle of 
incidence $\alpha_{\mbox{\tiny res}}\simeq 30^{\circ}$ according to the relation 
$\sin\alpha_{\mbox{\tiny res}}+\lambda/d=\sqrt{(1-n_e/n_c)/(2-n_e/n_c)}$ 
valid for a cold plasma, and assuming $n_e\gg n_c$  with 
$n_c=1.1 \times 10^{21}~\mbox{cm}^{-3}(\lambda/\mu\mbox{m})^{-2}$ 
the cut-off density.
The angle of incidence $\alpha$ was changed by pivoting the target holder 
around its vertical axis.

Proton spectra were recorded with a Thomson Parabola (TP) inside the vacuum chamber, equipped with a $100~\mu\mbox{m}$ diameter entrance pinhole at a distance of 150~mm from the target chamber center (TCC). Detection was provided by a micro channel plate plus a phosphor screen imaged onto a 12 bit charged coupled device (CCD) camera. 
For each value of $\alpha$ the TP was moved around the TCC and realigned with the normal to the rear surface of the target.  
(see Fig.\ref{setup_h}). 
The reflected laser light was imaged on a frosted glass placed at about 200~mm from TCC and recorded by a 12~bit CCD, in order to estimate variations in the target reflectivity.  
An optical fiber spectrometer was used to simultaneously record second harmonic ($2\omega$) and three-halves harmonic ($3\omega/2$) signals. 
Finally, a radiochromic film (RCF) stack was arranged in order to form a 50~mm diameter retractable ring around the target (Fig.\ref{setup_h}) and collect the particle and radiation emission over an angle of almost $2\pi$ radians (a 30$^{\circ}$ window was left open for laser entrance). The stack was composed by three HD-810 Gafcromic film layers, screened from visible and low-energy X-ray emission by a 2 $\mu$m aluminized Mylar film, and mostly sensitive to protons with energy of 2.5, 3.75 and 5~MeV, respectively.

\begin{figure}[t]
\includegraphics[width=0.48\textwidth]{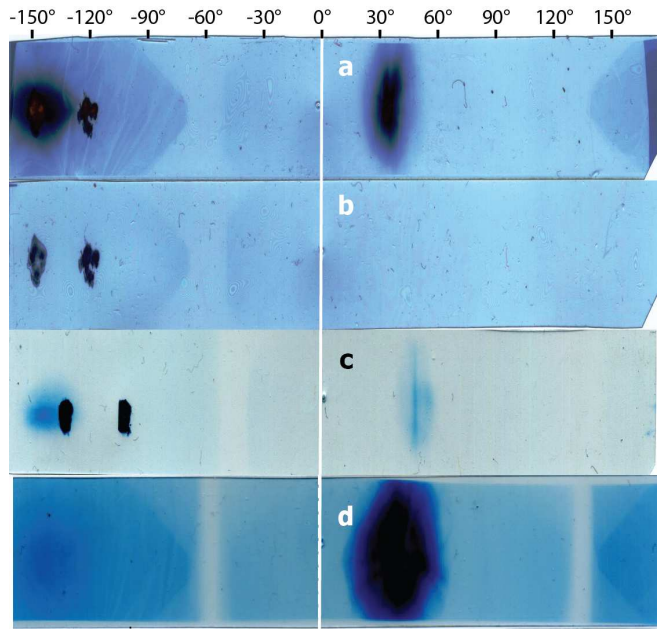}
\caption{Frames~\textsf{a} and~\textsf{b}: the first two layers of the exposed RCF at high laser contrast conditions, showing the typical proton forward (30$^{\circ}$) and backward (-150$^{\circ}$) proton spots as well as two smaller spots ($\simeq$-125$^{\circ}$ and $\simeq$-155$^{\circ}$) attributed to laser reflection at 0 and +1 grating diffraction orders. The position 0$^{\circ}$ corresponds to the incident laser axis. 
The parabolic ``shadows'' are due to the boundary of the target holder which screens diffuse radiation (such as electrons and hard X-rays) from the plasma.
Frame~\textsf{c}: same as \textsf{a} but for a angle of incidence different by 15$^{\circ}$, leading to a shift of the diffraction spots.
Frame~\textsf{d}: first RCF layer in low laser contrast conditions, for which both the backward proton and the diffraction spots disappear. 
All sets of RCF have been exposed to three laser shots.
\label{RCF}}
\end{figure}

A confirmation that the grating pattern is preserved until the interaction 
was provided by the RCF stack. 
Fig.\ref{RCF}~\textsf{a} and~\textsf{b} 
show the first two layers of the RCF stack after a shot on a GT with HC pulses at 30$^{\circ}$ incidence angle, that corresponds to 0$^{\circ}$ on the axis in the figure. 
Besides the expected proton spots in the  forward (30$^{\circ}$) and backward (-150$^{\circ}$) target normal directions (which is a typical feature of HC conditions \cite{ceccottiPRL07})
on the first stack film, two burn spots 
were obtained on the Al cover foil, corresponding to the two small structures at $\simeq$~-125$^{\circ}$ and $\simeq$~-155$^{\circ}$ in Fig.\ref{RCF}~\textsf{a}-\textsf{b}.
According to their angular positions and because of the reflected laser intensity on the RCF ($\sim 10^{14}~\mbox{W/cm}^{2}$) the spots can be attributed to reflection at the 0 and +1 grating diffraction orders. The spots are still present (with the expected angular shift) at an incidence angle to 45$^{\circ}$ (Fig.\ref{RCF}~\textsf{c}) but disappear under low  contrast ($10^{8}$) conditions (Fig.\ref{RCF}~\textsf{d}). 
The presence of an underdense preplasma
was also ruled out by the absence of optical emission at the $3\omega/2$ frequency, since the latter is 
tightly correlated to the plasma scalelength at the $n_e=n_{c}/4$ layer, where the underlying process of two-plasmon decay occurs
\cite{Veisz2002,*Tarasevitch2003,*Gizzi2007}.

\begin{figure}[t]
\includegraphics[width=0.48\textwidth]{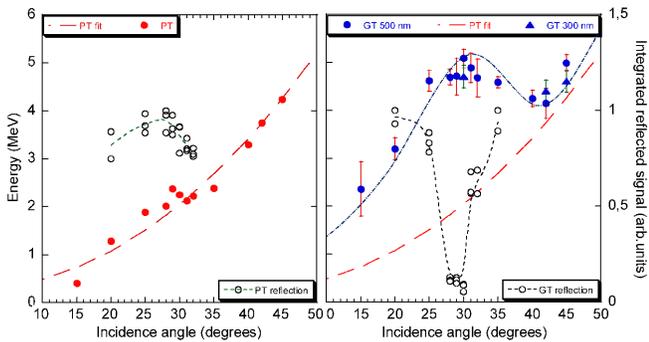}
\caption{Maximum proton energy (filled data points) and reflected light signal (empty data points) as a function of incidence angle $\alpha$. Left and right frames correspond to 20~$\mu\mbox{m}$ thick plane targets and to 23~$\mu\mbox{m}$ thick grating targets, respectively. Filled circles and triangles correspond to 0.5~$\mu$m and 0.3~$\mu$m deep gratings, respectively. The (red) dashed line in is proportional to $\sin^2\alpha/\cos\alpha$. The other lines are guides for the eye.
\label{angular_scan}}
\end{figure}

The maximum proton energy was measured for both GT and plane targets (PT) 
as a function of $\alpha$. 
The results for 20~$\mu\mbox{m}$ thick plane targets and 23~$\mu\mbox{m}$ thick GT are shown in Fig.\ref{angular_scan}. Most of the data have been obtained for a grating depth of 0.5~$\mu$m, with a few points from 0.3~$\mu$m deep GT yielding very similar energy values.
The PT show a variation of proton energy with $\alpha$ which is well fitted by a $\sin^2\alpha/\cos\alpha$ function. Such scaling may be simply understood as due to the variation of the normal component of the electric field ${\bf E}$ 
($\propto\sin\alpha$) and of the focal spot size ($\propto 1/\cos\alpha$) \cite{ceccottiPRL07}. 
In contrast, for GT the proton energy has a broad maximum (corresponding to $\approx$2.5 times the plane target energy for the same angle) near the resonant angle of 30$^{\circ}$. 
Fig.\ref{angular_scan} suggests that the resonant peak overlaps to the ``geometrical'' $\sin^2\alpha/\cos\alpha$ scaling, thus the energy might be increased using gratings with larger resonant angles. 
The reflected light signal from the frosted glass (also reported in Fig.\ref{angular_scan}) shows a dip around 30$^{\circ}$ for the grating targets, which is not observed for PT and is a signature of increased absorption.

A significant enhancement of the proton energy is also observed for small incidence angles (down to 15$^{\circ}$), far from the resonant value. 
This effect is explained by a mechanism similar to that observed in 
targets covered with regular pattern of microspheres
\cite{margaronePRL12,*floquetJAP13}. At a structured surface, electrons can be
dragged out in vacuum from the tip of a modulation by the component of 
${\bf E}$ parallel to the target plane even at normal incidence,
and may re-enter into the plasma near the tip of a neighboring modulation
there delivering their energy, similar to the simple model of
``vacuum heating'' absorption \cite{brunelPRL87} that thus becomes efficient 
also at small angles.
For large angles of incidence and $P$-polarization the electron 
motion near the laser-plasma interface is dominated by the 
component of ${\bf E}$ perpendicular to the surface, thus the structured 
targets behave more similarly to the plane ones.
Using \textit{S}-polarization, for which both ``vacuum heating'' of electrons by ${\bf E}$ and SW excitation in GT are ruled out, no protons of energy above the detection threshold of $\simeq 400~\mbox{keV}$ were observed for both PT and GT. This suggests that ``${\bf J}\times{\bf B}$'' heating effects are negligible due to the high plasma density, despite the relativistically strong intensity.

\begin{figure}
\includegraphics[width=0.48\textwidth]{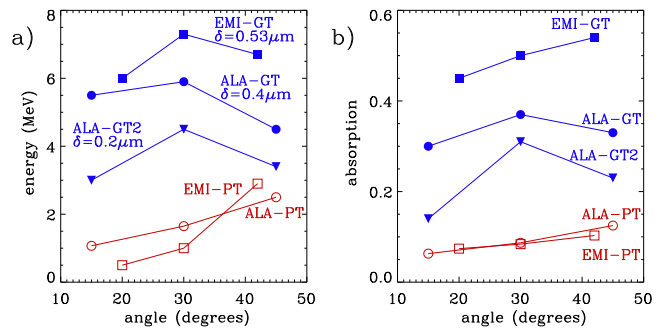}
\caption{2D PIC simulation results: 
a) cut-off energy of protons; b): fractional absorption.
Empty and filled symbols refer to plane 
targets (PT) and grating targets (GT), respectively.
Squares are data from EMI2D (EMI) simulations with grating depth 
$\delta=0.53~\mu\mbox{m}$ for GT, and circles and triangles are data from 
ALaDyn (ALA) simulations for $\delta=0.4~\mu\mbox{m}$ and $0.2~\mu\mbox{m}$,
respectively.
The data correspond to a time $t=350$~fs for EMI2D and $t=200$~fs for ALaDyn, respectively, relative to the time $t=0$ at which the pulse peak reaches the target.
\label{fig:pic1}}
\end{figure}

Two simulation campaigns using different PIC codes, EMI2D \cite{bigongiariPoP11} and ALaDyn \cite{benedettiIEEE08} have been performed to support the interpretation of the experimental results. The EMI2D simulations considered a $20~\mu\mbox{m}$ thick, proton plasma slab with density $n_e=100n_c=1.56 \times 10^{23}~\mbox{cm}^{-3}$ and initial temperatures $T_e=10~\mbox{keV}$, $T_i=1~\mbox{keV}$, 
and a laser pulse of $30~\mbox{fs}$ duration (FWHM of Gaussian envelope), 
$1.6\times 10^{19}~\mbox{W/cm}^2$ peak intensity, and homogeneous in the transverse direction (plane wave).  
The ALaDyn simulations considered a $n_e=120n_c=1.87 \times 10^{23}~\mbox{cm}^{-3}$, $T_e=T_i=0$, two-species slab composed of a $0.8~\mu\mbox{m}$ thick $Z/A=1/2$ layer a $0.05\mu\mbox{m}$ thick rear layer of protons, and a laser pulse with $25~\mbox{fs}$ duration (FWHM of $\sin^2$ envelope), $2\times 10^{19}~\mbox{W/cm}^2$ peak intensity, and a Gaussian transverse profile with $4~\mu\mbox{m}$ focal waist diameter. 
The grating periodicity was $2\lambda=1.6~\mu\mbox{m}$ in both cases 
and as in the experiment and 
different values of the peak-to-valley grating depth $\delta$ were investigated 
($\delta=0.53~\mu\mbox{m}$ for EMI2D and $\delta=0.2-0.4~\mu\mbox{m}$ 
for ALaDyn, respectively).

Figure~\ref{fig:pic1} shows simulation results for the maximum energy of protons and the fractional absorption, as a function of the incidence angle. 
A quantitative comparison with experimental data is not implied because of the unavoidable computational limitations in the PIC modeling of TNSA  and the reduction to a 2D geometry (see e.g. \cite{sgattoniPRE12}).
Nevertheless, the qualitative behavior observed in the experiment is reproduced by both sets of simulations, the maximum energy being in correspondance to the resonant angle.
The agreement is improved for the smallest value of the grating
depth ($\delta=0.2~\mu\mbox{m}$), 
about half the nominal value; in this case, the
cut-off energy for GTs is close to the plane targets value at 45$^{\circ}$, and
the enhancement factor at 15$^{\circ}$ also gets closer to the experimental
result. This observation suggests that some smoothing of the surface modulation
occurs during the interaction or prior to it, as for instance a residual 
picosecond pedestal to the femtosecond pulse may lead to some pre-heating 
and expansion of the target. A smaller depth of the grating reduces the 
``geometrical enhancement'' out of resonance while does not prevent the latter
to occur since the periodicity is preserved. 
Additional EMI2D simulations confirmed that the energy enhancement in GT does neither disappear in the presence of a sub-$\lambda$ density gradient at the front side nor for longer 60~fs pulses 
(which favor expansion and smoothing during the interaction)
allowing to obtain energies up to 18~MeV \cite{bigongiari-thesis}.

For both codes, additional simulations showed that the resonance smears out by either decreasing the plasma density or increasing the laser intensity. This observation further supports the evidence that the interaction occurs at solid density and that at high intensity there is a significant SW resonance broadening
due either to  ``detuning'' of the plasma frequency, which depends on the field amplitude in the relativistic regime, or to strong absorption. The latter is 
strongly increased in grating targets simulations, although not strictly proportional to the proton energy 
as shown in Fig.\ref{fig:pic1}~b).

\begin{figure}[t]
\includegraphics[width=0.48\textwidth]{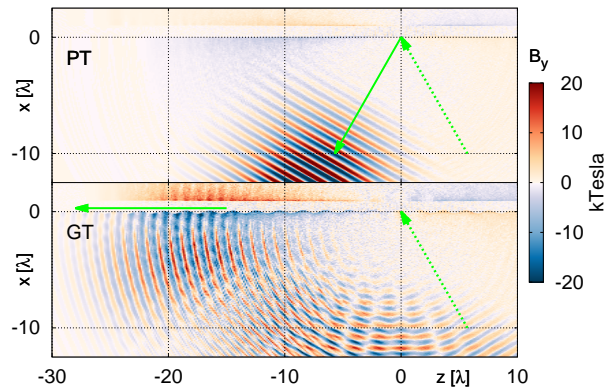}
\caption{2D PIC simulation results : snapshots at $t=75$~fs of the magnetic field $B_y$ (normal to the simulation plane) in the interaction at 30$^{\circ}$ with both a plane (PT) and grating (GT) target of $0.8~\mu\mbox{m}$ thickness. The laser axis of incidence is marked by the dashed arrows. For the GT, a wave propagating near to the target surface (on the left upper side) and reflection at several diffraction orders are apparent, while the PT plot is dominated by the specularly reflected pulse. The thick arrows give the propagation direction of the surface (for GT) and reflected (for PT) waves. See text for parameters. 
\label{fig:pic}}
\end{figure}

Figure~\ref{fig:pic} (from ALaDyn thin foil simulations) shows the magnetic field component $B_y$ normal to the simulation plane in the case of incidence 30$^{\circ}$, for both a PT and a resonant GT. In the latter case, the strong component localized near the surface and propagating in the direction of incidence is a signature of the SW excitation. 
The phase velocity $v_f$ and wavelength are very close to $c$ and $\lambda$, respectively, in agreement with the SW dispersion relation that gives $v_f/c \simeq 1-n_c/(2n_e)=0.996$ for $n_e/n_c=120$.  
The comparison also shows the lower amplitude of the field reflected from the GT as well as the reflection at several diffraction orders.

In conclusion, we have provided experimental evidence of absorption enhancement which is consistent with the resonant excitation of a surface wave in a grating target, at laser intensities higher than $10^{19}~\mbox{W cm}^{-2}$. 
The increase in coupling efficiency has been observed most clearly through the measurement of maximum proton energies emitted from the target. 
To further increase the proton energy either different values of the grating periodicity and the target thickness may be used, or the grating modulation might be embedded in complex target designs (see e.g. \cite{gaillardPoP11}).
Our results show that the availability of laser system with ultra-high contrast may allow to use structured targets for enhanced absorption also at the highest intensities available today, and to extend investigations of plasmonics in the relativistic regime.

The authors acknowledge fruitful discussions with Dr. P.~Audebert (LULI, France). 
The research leading to these results has received funding from LASERLAB-EUROPE (grant no. 284464, EU's 7$^{\mbox{\tiny th}}$ FP), proposal n.SLIC001693. 
This work was also supported by the Conseil G\'{e}n\'{e}ral de l'Essonne (ASTRE program), by Region Ile de France (SESAME Project), 
by Saphir Consortium (OSEO), by RTRA Triangle de la Physique, by ANR (France) under ref. BLAN08-1\_380251, 
by Czech Science Foundation (project No. P205/11/1165), by ECOP project No. CZ.1.07/2.3.00/20.0087, 
by ELI-Italy funded by CNR,
and by MIUR Italy (FIR project SULDIS).
EMI2D simulations were performed using HPC resources from GENCI-CCRT (grant 2012-t2012056851). For ALaDyn simulations we acknowledge PRACE for access to resource FERMI based in Italy at CINECA, via the project LSAIL.


\begin{thebibliography}{45}%
\makeatletter
\providecommand \@ifxundefined [1]{%
 \@ifx{#1\undefined}
}%
\providecommand \@ifnum [1]{%
 \ifnum #1\expandafter \@firstoftwo
 \else \expandafter \@secondoftwo
 \fi
}%
\providecommand \@ifx [1]{%
 \ifx #1\expandafter \@firstoftwo
 \else \expandafter \@secondoftwo
 \fi
}%
\providecommand \natexlab [1]{#1}%
\providecommand \enquote  [1]{``#1''}%
\providecommand \bibnamefont  [1]{#1}%
\providecommand \bibfnamefont [1]{#1}%
\providecommand \citenamefont [1]{#1}%
\providecommand \href@noop [0]{\@secondoftwo}%
\providecommand \href [0]{\begingroup \@sanitize@url \@href}%
\providecommand \@href[1]{\@@startlink{#1}\@@href}%
\providecommand \@@href[1]{\endgroup#1\@@endlink}%
\providecommand \@sanitize@url [0]{\catcode `\\12\catcode `\$12\catcode
  `\&12\catcode `\#12\catcode `\^12\catcode `\_12\catcode `\%12\relax}%
\providecommand \@@startlink[1]{}%
\providecommand \@@endlink[0]{}%
\providecommand \url  [0]{\begingroup\@sanitize@url \@url }%
\providecommand \@url [1]{\endgroup\@href {#1}{\urlprefix }}%
\providecommand \urlprefix  [0]{URL }%
\providecommand \Eprint [0]{\href }%
\providecommand \doibase [0]{http://dx.doi.org/}%
\providecommand \selectlanguage [0]{\@gobble}%
\providecommand \bibinfo  [0]{\@secondoftwo}%
\providecommand \bibfield  [0]{\@secondoftwo}%
\providecommand \translation [1]{[#1]}%
\providecommand \BibitemOpen [0]{}%
\providecommand \bibitemStop [0]{}%
\providecommand \bibitemNoStop [0]{.\EOS\space}%
\providecommand \EOS [0]{\spacefactor3000\relax}%
\providecommand \BibitemShut  [1]{\csname bibitem#1\endcsname}%
\let\auto@bib@innerbib\@empty
\bibitem [{\citenamefont {Borghesi}\ \emph {et~al.}(2006)\citenamefont
  {Borghesi} \emph {et~al.}}]{borghesiFST06}%
  \BibitemOpen
  \bibfield  {author} {\bibinfo {author} {\bibfnamefont {M.}~\bibnamefont
  {Borghesi}} \emph {et~al.},\ }\href
  {http://www.ans.org/pubs/journals/fst/va-49-3-412-439} {\bibfield  {journal}
  {\bibinfo  {journal} {Fus. Sci. Techn.}\ }\textbf {\bibinfo {volume} {49}},\
  \bibinfo {pages} {412} (\bibinfo {year} {2006})}\BibitemShut {NoStop}%
\bibitem [{\citenamefont {Daido}\ \emph {et~al.}(2012)\citenamefont {Daido},
  \citenamefont {Nishiuchi},\ and\ \citenamefont {Pirozhkov}}]{daidoRPP12}%
  \BibitemOpen
  \bibfield  {author} {\bibinfo {author} {\bibfnamefont {H.}~\bibnamefont
  {Daido}}, \bibinfo {author} {\bibfnamefont {M.}~\bibnamefont {Nishiuchi}}, \
  and\ \bibinfo {author} {\bibfnamefont {A.~S.}\ \bibnamefont {Pirozhkov}},\
  }\href {\doibase 10.1088/0034-4885/75/5/056401} {\bibfield  {journal}
  {\bibinfo  {journal} {Rep. Prog. Phys.}\ }\textbf {\bibinfo {volume} {75}},\
  \bibinfo {pages} {056401} (\bibinfo {year} {2012})}\BibitemShut {NoStop}%
\bibitem [{\citenamefont {Macchi}\ \emph {et~al.}(2013)\citenamefont {Macchi},
  \citenamefont {Borghesi},\ and\ \citenamefont {Passoni}}]{macchiRMP13}%
  \BibitemOpen
  \bibfield  {author} {\bibinfo {author} {\bibfnamefont {A.}~\bibnamefont
  {Macchi}}, \bibinfo {author} {\bibfnamefont {M.}~\bibnamefont {Borghesi}}, \
  and\ \bibinfo {author} {\bibfnamefont {M.}~\bibnamefont {Passoni}},\ }\href
  {\doibase 10.1103/RevModPhys.85.751} {\bibfield  {journal} {\bibinfo
  {journal} {Rev. Mod. Phys.}\ }\textbf {\bibinfo {volume} {85}},\ \bibinfo
  {pages} {751} (\bibinfo {year} {2013})}\BibitemShut {NoStop}%
\bibitem [{\citenamefont {Toth}\ \emph {et~al.}(2007)\citenamefont {Toth} \emph
  {et~al.}}]{tothPoP07}%
  \BibitemOpen
  \bibfield  {author} {\bibinfo {author} {\bibfnamefont {R.}~\bibnamefont
  {Toth}} \emph {et~al.},\ }\href {\doibase 10.1063/1.2730778} {\bibfield
  {journal} {\bibinfo  {journal} {Phys. Plasmas}\ }\textbf {\bibinfo {volume}
  {14}},\ \bibinfo {eid} {053506} (\bibinfo {year} {2007})}\BibitemShut
  {NoStop}%
\bibitem [{\citenamefont {Chen}\ \emph {et~al.}(2008)\citenamefont {Chen} \emph
  {et~al.}}]{chenPRL08}%
  \BibitemOpen
  \bibfield  {author} {\bibinfo {author} {\bibfnamefont {L.~M.}\ \bibnamefont
  {Chen}} \emph {et~al.},\ }\href {\doibase 10.1103/PhysRevLett.100.045004}
  {\bibfield  {journal} {\bibinfo  {journal} {Phys. Rev. Lett.}\ }\textbf
  {\bibinfo {volume} {100}},\ \bibinfo {pages} {045004} (\bibinfo {year}
  {2008})}\BibitemShut {NoStop}%
\bibitem [{\citenamefont {{A. L\'{e}vy}}\ \emph {et~al.}(2010)\citenamefont
  {{A. L\'{e}vy}} \emph {et~al.}}]{levyAPL10}%
  \BibitemOpen
  \bibfield  {author} {\bibinfo {author} {\bibnamefont {{A. L\'{e}vy}}} \emph
  {et~al.},\ }\href {\doibase 10.1063/1.3386534} {\bibfield  {journal}
  {\bibinfo  {journal} {Appl. Phys. Lett.}\ }\textbf {\bibinfo {volume} {96}},\
  \bibinfo {eid} {151114} (\bibinfo {year} {2010})}\BibitemShut {NoStop}%
\bibitem [{\citenamefont {Zhang}\ \emph {et~al.}(2011)\citenamefont {Zhang}
  \emph {et~al.}}]{zhangOE11}%
  \BibitemOpen
  \bibfield  {author} {\bibinfo {author} {\bibfnamefont {Z.}~\bibnamefont
  {Zhang}} \emph {et~al.},\ }\href {\doibase 10.1364/OE.19.004560} {\bibfield
  {journal} {\bibinfo  {journal} {Opt. Express}\ }\textbf {\bibinfo {volume}
  {19}},\ \bibinfo {pages} {4560} (\bibinfo {year} {2011})}\BibitemShut
  {NoStop}%
\bibitem [{\citenamefont {Dobosz}\ \emph {et~al.}(2005)\citenamefont {Dobosz}
  \emph {et~al.}}]{doboszPRL05}%
  \BibitemOpen
  \bibfield  {author} {\bibinfo {author} {\bibfnamefont {S.}~\bibnamefont
  {Dobosz}} \emph {et~al.},\ }\href {\doibase 10.1103/PhysRevLett.95.025001}
  {\bibfield  {journal} {\bibinfo  {journal} {Phys. Rev. Lett.}\ }\textbf
  {\bibinfo {volume} {95}},\ \bibinfo {pages} {025001} (\bibinfo {year}
  {2005})}\BibitemShut {NoStop}%
\bibitem [{\citenamefont {Osterholz}\ \emph {et~al.}(2006)\citenamefont
  {Osterholz} \emph {et~al.}}]{osterholzPRL06}%
  \BibitemOpen
  \bibfield  {author} {\bibinfo {author} {\bibfnamefont {J.}~\bibnamefont
  {Osterholz}} \emph {et~al.},\ }\href {\doibase 10.1103/PhysRevLett.96.085002}
  {\bibfield  {journal} {\bibinfo  {journal} {Phys. Rev. Lett.}\ }\textbf
  {\bibinfo {volume} {96}},\ \bibinfo {pages} {085002} (\bibinfo {year}
  {2006})}\BibitemShut {NoStop}%
\bibitem [{\citenamefont {Perez}\ \emph {et~al.}(2010)\citenamefont {Perez}
  \emph {et~al.}}]{perezPRL10}%
  \BibitemOpen
  \bibfield  {author} {\bibinfo {author} {\bibfnamefont {F.}~\bibnamefont
  {Perez}} \emph {et~al.},\ }\href {\doibase 10.1103/PhysRevLett.104.085001}
  {\bibfield  {journal} {\bibinfo  {journal} {Phys. Rev. Lett.}\ }\textbf
  {\bibinfo {volume} {104}},\ \bibinfo {pages} {085001} (\bibinfo {year}
  {2010})}\BibitemShut {NoStop}%
\bibitem [{\citenamefont {Santos}\ \emph {et~al.}(2002)\citenamefont {Santos}
  \emph {et~al.}}]{santosPRL02}%
  \BibitemOpen
  \bibfield  {author} {\bibinfo {author} {\bibfnamefont {J.~J.}\ \bibnamefont
  {Santos}} \emph {et~al.},\ }\href {\doibase 10.1103/PhysRevLett.89.025001}
  {\bibfield  {journal} {\bibinfo  {journal} {Phys. Rev. Lett.}\ }\textbf
  {\bibinfo {volume} {89}},\ \bibinfo {pages} {025001} (\bibinfo {year}
  {2002})}\BibitemShut {NoStop}%
\bibitem [{\citenamefont {{P. K\"oster}}\ \emph {et~al.}(2009)\citenamefont
  {{P. K\"oster}} \emph {et~al.}}]{kosterPPCF09}%
  \BibitemOpen
  \bibfield  {author} {\bibinfo {author} {\bibnamefont {{P. K\"oster}}} \emph
  {et~al.},\ }\href {\doibase 10.1088/0741-3335/51/1/014007} {\bibfield
  {journal} {\bibinfo  {journal} {Plasma Phys. Contr. Fusion}\ }\textbf
  {\bibinfo {volume} {51}},\ \bibinfo {pages} {014007} (\bibinfo {year}
  {2009})}\BibitemShut {NoStop}%
\bibitem [{\citenamefont {McKenna}\ \emph {et~al.}(2011)\citenamefont {McKenna}
  \emph {et~al.}}]{mckennaPRL11}%
  \BibitemOpen
  \bibfield  {author} {\bibinfo {author} {\bibfnamefont {P.}~\bibnamefont
  {McKenna}} \emph {et~al.},\ }\href {\doibase 10.1103/PhysRevLett.106.185004}
  {\bibfield  {journal} {\bibinfo  {journal} {Phys. Rev. Lett.}\ }\textbf
  {\bibinfo {volume} {106}},\ \bibinfo {pages} {185004} (\bibinfo {year}
  {2011})}\BibitemShut {NoStop}%
\bibitem [{\citenamefont {Kulcs\'ar}\ \emph {et~al.}(2000)\citenamefont
  {Kulcs\'ar} \emph {et~al.}}]{kulcsarPRL00}%
  \BibitemOpen
  \bibfield  {author} {\bibinfo {author} {\bibfnamefont {G.}~\bibnamefont
  {Kulcs\'ar}} \emph {et~al.},\ }\href {\doibase 10.1103/PhysRevLett.84.5149}
  {\bibfield  {journal} {\bibinfo  {journal} {Phys. Rev. Lett.}\ }\textbf
  {\bibinfo {volume} {84}},\ \bibinfo {pages} {5149} (\bibinfo {year}
  {2000})}\BibitemShut {NoStop}%
\bibitem [{\citenamefont {Bagchi}\ \emph {et~al.}(2007)\citenamefont {Bagchi}
  \emph {et~al.}}]{bagchiAPB07}%
  \BibitemOpen
  \bibfield  {author} {\bibinfo {author} {\bibfnamefont {S.}~\bibnamefont
  {Bagchi}} \emph {et~al.},\ }\href {\doibase 10.1007/s00340-007-2706-7}
  {\bibfield  {journal} {\bibinfo  {journal} {Appl. Phys. B: Lasers Opt.}\
  }\textbf {\bibinfo {volume} {88}},\ \bibinfo {pages} {167} (\bibinfo {year}
  {2007})}\BibitemShut {NoStop}%
\bibitem [{\citenamefont {Sumeruk}\ \emph {et~al.}(2007)\citenamefont {Sumeruk}
  \emph {et~al.}}]{sumerukPRL07}%
  \BibitemOpen
  \bibfield  {author} {\bibinfo {author} {\bibfnamefont {H.~A.}\ \bibnamefont
  {Sumeruk}} \emph {et~al.},\ }\href {\doibase 10.1103/PhysRevLett.98.045001}
  {\bibfield  {journal} {\bibinfo  {journal} {Phys. Rev. Lett.}\ }\textbf
  {\bibinfo {volume} {98}},\ \bibinfo {pages} {045001} (\bibinfo {year}
  {2007})}\BibitemShut {NoStop}%
\bibitem [{\citenamefont {Kneip}\ \emph {et~al.}(2008)\citenamefont {Kneip}
  \emph {et~al.}}]{kneipHEDP08}%
  \BibitemOpen
  \bibfield  {author} {\bibinfo {author} {\bibfnamefont {S.}~\bibnamefont
  {Kneip}} \emph {et~al.},\ }\href {\doibase 10.1016/j.hedp.2007.10.002}
  {\bibfield  {journal} {\bibinfo  {journal} {High Energy Density Phys.}\
  }\textbf {\bibinfo {volume} {4}},\ \bibinfo {pages} {41 } (\bibinfo {year}
  {2008})}\BibitemShut {NoStop}%
\bibitem [{\citenamefont {Chakravarty}\ \emph {et~al.}(2011)\citenamefont
  {Chakravarty} \emph {et~al.}}]{chakravartyAPB11}%
  \BibitemOpen
  \bibfield  {author} {\bibinfo {author} {\bibfnamefont {U.}~\bibnamefont
  {Chakravarty}} \emph {et~al.},\ }\href {\doibase 10.1007/s00340-011-4434-2}
  {\bibfield  {journal} {\bibinfo  {journal} {Appl. Phys. B: Lasers Opt.}\
  }\textbf {\bibinfo {volume} {103}},\ \bibinfo {pages} {571} (\bibinfo {year}
  {2011})}\BibitemShut {NoStop}%
\bibitem [{\citenamefont {Mondal}\ \emph {et~al.}(2011)\citenamefont {Mondal}
  \emph {et~al.}}]{mondalPRB11}%
  \BibitemOpen
  \bibfield  {author} {\bibinfo {author} {\bibfnamefont {S.}~\bibnamefont
  {Mondal}} \emph {et~al.},\ }\href {\doibase 10.1103/PhysRevB.83.035408}
  {\bibfield  {journal} {\bibinfo  {journal} {Phys. Rev. B}\ }\textbf {\bibinfo
  {volume} {83}},\ \bibinfo {pages} {035408} (\bibinfo {year}
  {2011})}\BibitemShut {NoStop}%
\bibitem [{\citenamefont {Zigler}\ \emph {et~al.}(2011)\citenamefont {Zigler}
  \emph {et~al.}}]{ziglerPRL11}%
  \BibitemOpen
  \bibfield  {author} {\bibinfo {author} {\bibfnamefont {A.}~\bibnamefont
  {Zigler}} \emph {et~al.},\ }\href {\doibase 10.1103/PhysRevLett.106.134801}
  {\bibfield  {journal} {\bibinfo  {journal} {Phys. Rev. Lett.}\ }\textbf
  {\bibinfo {volume} {106}},\ \bibinfo {pages} {134801} (\bibinfo {year}
  {2011})}\BibitemShut {NoStop}%
\bibitem [{\citenamefont {Raether}(1988)}]{raether-book}%
  \BibitemOpen
  \bibfield  {author} {\bibinfo {author} {\bibfnamefont {H.}~\bibnamefont
  {Raether}},\ }
  {\emph {\bibinfo {title} {Surface plasmons on smooth and rough surfaces and on
  gratings}}},\ \bibinfo {series} {Springer tracts in modern physics}\ No.\
  \bibinfo {number} {111}\ (\bibinfo  {publisher} {Springer},\ \bibinfo {year}
  {1988})\BibitemShut {NoStop}%
\bibitem [{\citenamefont {Maier}\ and\ \citenamefont
  {Atwater}(2005)}]{maierJAP05}%
  \BibitemOpen
  \bibfield  {author} {\bibinfo {author} {\bibfnamefont {S.~A.}\ \bibnamefont
  {Maier}}\ and\ \bibinfo {author} {\bibfnamefont {H.~A.}\ \bibnamefont
  {Atwater}},\ }\href {\doibase 10.1063/1.1951057} {\bibfield  {journal}
  {\bibinfo  {journal} {J. Appl. Phys.}\ }\textbf {\bibinfo {volume} {98}},\
  \bibinfo {eid} {011101} (\bibinfo {year} {2005})}\BibitemShut {NoStop}%
\bibitem [{\citenamefont {Gauthier}\ \emph {et~al.}(1995)\citenamefont
  {Gauthier} \emph {et~al.}}]{gauthierSPIE95}%
  \BibitemOpen
  \bibfield  {author} {\bibinfo {author} {\bibfnamefont {J.-C.}\ \bibnamefont
  {Gauthier}} \emph {et~al.},\ }\href {\doibase 10.1117/12.220985} {\bibfield
  {journal} {\bibinfo  {journal} {Proc. SPIE}\ }\textbf {\bibinfo {volume}
  {2523}},\ \bibinfo {pages} {242} (\bibinfo {year} {1995})}\BibitemShut
  {NoStop}%
\bibitem [{\citenamefont {Kahaly}\ \emph {et~al.}(2008)\citenamefont {Kahaly}
  \emph {et~al.}}]{kahalyPRL08}%
  \BibitemOpen
  \bibfield  {author} {\bibinfo {author} {\bibfnamefont {S.}~\bibnamefont
  {Kahaly}} \emph {et~al.},\ }\href {\doibase 10.1103/PhysRevLett.101.145001}
  {\bibfield  {journal} {\bibinfo  {journal} {Phys. Rev. Lett.}\ }\textbf
  {\bibinfo {volume} {101}},\ \bibinfo {pages} {145001} (\bibinfo {year}
  {2008})}\BibitemShut {NoStop}%
\bibitem [{\citenamefont {Hu}\ \emph {et~al.}(2010)\citenamefont {Hu} \emph
  {et~al.}}]{huPoP10}%
  \BibitemOpen
  \bibfield  {author} {\bibinfo {author} {\bibfnamefont {G.}~\bibnamefont {Hu}}
  \emph {et~al.},\ }\href {\doibase 10.1063/1.3368792} {\bibfield  {journal}
  {\bibinfo  {journal} {Phys. Plasmas}\ }\textbf {\bibinfo {volume} {17}},\
  \bibinfo {eid} {033109} (\bibinfo {year} {2010})}\BibitemShut {NoStop}%
\bibitem [{\citenamefont {Bagchi}\ \emph {et~al.}(2012)\citenamefont {Bagchi}
  \emph {et~al.}}]{bagchiPoP12}%
  \BibitemOpen
  \bibfield  {author} {\bibinfo {author} {\bibfnamefont {S.}~\bibnamefont
  {Bagchi}} \emph {et~al.},\ }\href {\doibase 10.1063/1.3693388} {\bibfield
  {journal} {\bibinfo  {journal} {Phys. Plasmas}\ }\textbf {\bibinfo {volume}
  {19}},\ \bibinfo {eid} {030703} (\bibinfo {year} {2012})}\BibitemShut
  {NoStop}%
\bibitem [{\citenamefont {Dromey}\ \emph {et~al.}(2004)\citenamefont {Dromey},
  \citenamefont {Kar}, \citenamefont {Zepf},\ and\ \citenamefont
  {Foster}}]{dromeyRSI04}%
  \BibitemOpen
  \bibfield  {author} {\bibinfo {author} {\bibfnamefont {B.}~\bibnamefont
  {Dromey}}, \bibinfo {author} {\bibfnamefont {S.}~\bibnamefont {Kar}},
  \bibinfo {author} {\bibfnamefont {M.}~\bibnamefont {Zepf}}, \ and\ \bibinfo
  {author} {\bibfnamefont {P.}~\bibnamefont {Foster}},\ }\href {\doibase
  10.1063/1.1646737} {\bibfield  {journal} {\bibinfo  {journal} {Rev. Sci.
  Instrum.}\ }\textbf {\bibinfo {volume} {75}},\ \bibinfo {pages} {645}
  (\bibinfo {year} {2004})}\BibitemShut {NoStop}%
\bibitem [{\citenamefont {Thaury}\ \emph {et~al.}(2007)\citenamefont {Thaury}
  \emph {et~al.}}]{thauryNP07}%
  \BibitemOpen
  \bibfield  {author} {\bibinfo {author} {\bibfnamefont {C.}~\bibnamefont
  {Thaury}} \emph {et~al.},\ }\href {\doibase 10.1038/nphys595} {\bibfield
  {journal} {\bibinfo  {journal} {Nature Phys.}\ }\textbf {\bibinfo {volume}
  {3}},\ \bibinfo {pages} {424} (\bibinfo {year} {2007})}\BibitemShut {NoStop}%
\bibitem [{\citenamefont {Cerchez}\ \emph {et~al.}(2013)\citenamefont {Cerchez}
  \emph {et~al.}}]{cerchezPRL13}%
  \BibitemOpen
  \bibfield  {author} {\bibinfo {author} {\bibfnamefont {M.}~\bibnamefont
  {Cerchez}} \emph {et~al.},\ }\href {\doibase 10.1103/PhysRevLett.110.065003}
  {\bibfield  {journal} {\bibinfo  {journal} {Phys. Rev. Lett.}\ }\textbf
  {\bibinfo {volume} {110}},\ \bibinfo {pages} {065003} (\bibinfo {year}
  {2013})}\BibitemShut {NoStop}%
\bibitem [{\citenamefont {Raynaud}\ \emph {et~al.}(2007)\citenamefont
  {Raynaud}, \citenamefont {Kupersztych}, \citenamefont {Riconda},
  \citenamefont {Adam},\ and\ \citenamefont {Heron}}]{raynaudPoP07}%
  \BibitemOpen
  \bibfield  {author} {\bibinfo {author} {\bibfnamefont {M.}~\bibnamefont
  {Raynaud}}, \bibinfo {author} {\bibfnamefont {J.}~\bibnamefont
  {Kupersztych}}, \bibinfo {author} {\bibfnamefont {C.}~\bibnamefont
  {Riconda}}, \bibinfo {author} {\bibfnamefont {J.~C.}\ \bibnamefont {Adam}}, \
  and\ \bibinfo {author} {\bibfnamefont {A.}~\bibnamefont {Heron}},\ }\href
  {\doibase 10.1063/1.2755969} {\bibfield  {journal} {\bibinfo  {journal}
  {Phys. Plasmas}\ }\textbf {\bibinfo {volume} {14}},\ \bibinfo {eid} {092702}
  (\bibinfo {year} {2007})}\BibitemShut {NoStop}%
\bibitem [{\citenamefont {Bigongiari}\ \emph {et~al.}(2011)\citenamefont
  {Bigongiari}, \citenamefont {Raynaud}, \citenamefont {Riconda}, \citenamefont
  {H\'{e}ron},\ and\ \citenamefont {Macchi}}]{bigongiariPoP11}%
  \BibitemOpen
  \bibfield  {author} {\bibinfo {author} {\bibfnamefont {A.}~\bibnamefont
  {Bigongiari}}, \bibinfo {author} {\bibfnamefont {M.}~\bibnamefont {Raynaud}},
  \bibinfo {author} {\bibfnamefont {C.}~\bibnamefont {Riconda}}, \bibinfo
  {author} {\bibfnamefont {A.}~\bibnamefont {H\'{e}ron}}, \ and\ \bibinfo
  {author} {\bibfnamefont {A.}~\bibnamefont {Macchi}},\ }\href {\doibase
  10.1063/1.3646520} {\bibfield  {journal} {\bibinfo  {journal} {Phys.
  Plasmas}\ }\textbf {\bibinfo {volume} {18}},\ \bibinfo {eid} {102701}
  (\bibinfo {year} {2011})}\BibitemShut {NoStop}%
\bibitem [{\citenamefont {Bigongiari}\ \emph {et~al.}(2013)\citenamefont
  {Bigongiari}, \citenamefont {Raynaud}, \citenamefont {Riconda},\ and\
  \citenamefont {H\'{e}ron}}]{bigongiariPoP13}%
  \BibitemOpen
  \bibfield  {author} {\bibinfo {author} {\bibfnamefont {A.}~\bibnamefont
  {Bigongiari}}, \bibinfo {author} {\bibfnamefont {M.}~\bibnamefont {Raynaud}},
  \bibinfo {author} {\bibfnamefont {C.}~\bibnamefont {Riconda}}, \ and\
  \bibinfo {author} {\bibfnamefont {A.}~\bibnamefont {H\'{e}ron}},\ }\href
  {\doibase 10.1063/1.4802989} {\bibfield  {journal} {\bibinfo  {journal}
  {Phys. Plasmas}\ }\textbf {\bibinfo {volume} {20}},\ \bibinfo {eid} {052701}
  (\bibinfo {year} {2013})}\BibitemShut {NoStop}%
\bibitem [{\citenamefont {L\'{e}vy}\ \emph {et~al.}(2007)\citenamefont
  {L\'{e}vy} \emph {et~al.}}]{levy1}%
  \BibitemOpen
  \bibfield  {author} {\bibinfo {author} {\bibfnamefont {A.}~\bibnamefont
  {L\'{e}vy}} \emph {et~al.},\ }\href {\doibase 10.1364/OL.32.000310}
  {\bibfield  {journal} {\bibinfo  {journal} {Opt. Lett.}\ }\textbf {\bibinfo
  {volume} {32}},\ \bibinfo {pages} {310} (\bibinfo {year} {2007})}\BibitemShut
  {NoStop}%
\bibitem [{\citenamefont {Kapteyn}\ \emph {et~al.}(1991)\citenamefont
  {Kapteyn}, \citenamefont {Murnane}, \citenamefont {Szoke},\ and\
  \citenamefont {Falcone}}]{kapteyn}%
  \BibitemOpen
  \bibfield  {author} {\bibinfo {author} {\bibfnamefont {H.~C.}\ \bibnamefont
  {Kapteyn}}, \bibinfo {author} {\bibfnamefont {M.~M.}\ \bibnamefont
  {Murnane}}, \bibinfo {author} {\bibfnamefont {A.}~\bibnamefont {Szoke}}, \
  and\ \bibinfo {author} {\bibfnamefont {R.~W.}\ \bibnamefont {Falcone}},\
  }\href {\doibase 10.1364/OL.16.000490} {\bibfield  {journal} {\bibinfo
  {journal} {Opt. Lett.}\ }\textbf {\bibinfo {volume} {16}},\ \bibinfo {pages}
  {490} (\bibinfo {year} {1991})}\BibitemShut {NoStop}%
\bibitem [{\citenamefont {Ceccotti}\ \emph {et~al.}(2007)\citenamefont
  {Ceccotti} \emph {et~al.}}]{ceccottiPRL07}%
  \BibitemOpen
  \bibfield  {author} {\bibinfo {author} {\bibfnamefont {T.}~\bibnamefont
  {Ceccotti}} \emph {et~al.},\ }\href {\doibase 10.1103/PhysRevLett.99.185002}
  {\bibfield  {journal} {\bibinfo  {journal} {Phys. Rev. Lett.}\ }\textbf
  {\bibinfo {volume} {99}},\ \bibinfo {pages} {185002} (\bibinfo {year}
  {2007})}\BibitemShut {NoStop}%
\bibitem [{\citenamefont {Veisz}\ \emph {et~al.}(2002)\citenamefont {Veisz}
  \emph {et~al.}}]{Veisz2002}%
  \BibitemOpen
  \bibfield  {author} {\bibinfo {author} {\bibfnamefont {L.}~\bibnamefont
  {Veisz}} \emph {et~al.},\ }\href {\doibase 10.1063/1.1493794} {\bibfield
  {journal} {\bibinfo  {journal} {Phys. Plasmas}\ }\textbf {\bibinfo {volume}
  {9}},\ \bibinfo {pages} {3197} (\bibinfo {year} {2002})}\BibitemShut
  {NoStop}%
\bibitem [{\citenamefont {Tarasevitch}\ \emph {et~al.}(2003)\citenamefont
  {Tarasevitch}, \citenamefont {Dietrich}, \citenamefont {Blome}, \citenamefont
  {Sokolowski-Tinten},\ and\ \citenamefont {{von der
  Linde}}}]{Tarasevitch2003}%
  \BibitemOpen
  \bibfield  {author} {\bibinfo {author} {\bibfnamefont {A.}~\bibnamefont
  {Tarasevitch}}, \bibinfo {author} {\bibfnamefont {C.}~\bibnamefont
  {Dietrich}}, \bibinfo {author} {\bibfnamefont {C.}~\bibnamefont {Blome}},
  \bibinfo {author} {\bibfnamefont {K.}~\bibnamefont {Sokolowski-Tinten}}, \
  and\ \bibinfo {author} {\bibfnamefont {D.}~\bibnamefont {{von der Linde}}},\
  }\href {\doibase 10.1103/PhysRevE.68.026410} {\bibfield  {journal} {\bibinfo
  {journal} {Phys. Rev. E}\ }\textbf {\bibinfo {volume} {68}},\ \bibinfo
  {pages} {026410} (\bibinfo {year} {2003})}\BibitemShut {NoStop}%
\bibitem [{\citenamefont {Gizzi}\ \emph {et~al.}(2007)\citenamefont {Gizzi}
  \emph {et~al.}}]{Gizzi2007}%
  \BibitemOpen
  \bibfield  {author} {\bibinfo {author} {\bibfnamefont {L.~A.}\ \bibnamefont
  {Gizzi}} \emph {et~al.},\ }\href {\doibase 10.1117/12.742103} {\bibfield
  {journal} {\bibinfo  {journal} {Proc. SPIE}\ }\textbf {\bibinfo {volume}
  {6634}},\ \bibinfo {pages} {66341H} (\bibinfo {year} {2007})}\BibitemShut
  {NoStop}%
\bibitem [{\citenamefont {Margarone}\ \emph {et~al.}(2012)\citenamefont
  {Margarone} \emph {et~al.}}]{margaronePRL12}%
  \BibitemOpen
  \bibfield  {author} {\bibinfo {author} {\bibfnamefont {D.}~\bibnamefont
  {Margarone}} \emph {et~al.},\ }\href {\doibase
  10.1103/PhysRevLett.109.234801} {\bibfield  {journal} {\bibinfo  {journal}
  {Phys. Rev. Lett.}\ }\textbf {\bibinfo {volume} {109}},\ \bibinfo {pages}
  {234801} (\bibinfo {year} {2012})}\BibitemShut {NoStop}%
\bibitem [{\citenamefont {Floquet}\ \emph {et~al.}(2013)\citenamefont {Floquet}
  \emph {et~al.}}]{floquetJAP13}%
  \BibitemOpen
  \bibfield  {author} {\bibinfo {author} {\bibfnamefont {V.}~\bibnamefont
  {Floquet}} \emph {et~al.},\ }\href {\doibase 10.1063/1.4819239} {\bibfield
  {journal} {\bibinfo  {journal} {J. Appl. Phys.}\ }\textbf {\bibinfo {volume}
  {114}},\ \bibinfo {eid} {083305} (\bibinfo {year} {2013})}\BibitemShut
  {NoStop}%
\bibitem [{\citenamefont {Brunel}(1987)}]{brunelPRL87}%
  \BibitemOpen
  \bibfield  {author} {\bibinfo {author} {\bibfnamefont {F.}~\bibnamefont
  {Brunel}},\ }\href {\doibase 10.1103/PhysRevLett.59.52} {\bibfield  {journal}
  {\bibinfo  {journal} {Phys. Rev. Lett.}\ }\textbf {\bibinfo {volume} {59}},\
  \bibinfo {pages} {52} (\bibinfo {year} {1987})}\BibitemShut {NoStop}%
\bibitem [{\citenamefont {Benedetti}\ \emph {et~al.}(2008)\citenamefont
  {Benedetti}, \citenamefont {Sgattoni}, \citenamefont {Turchetti},\ and\
  \citenamefont {Londrillo}}]{benedettiIEEE08}%
  \BibitemOpen
  \bibfield  {author} {\bibinfo {author} {\bibfnamefont {C.}~\bibnamefont
  {Benedetti}}, \bibinfo {author} {\bibfnamefont {A.}~\bibnamefont {Sgattoni}},
  \bibinfo {author} {\bibfnamefont {G.}~\bibnamefont {Turchetti}}, \ and\
  \bibinfo {author} {\bibfnamefont {P.}~\bibnamefont {Londrillo}},\ }\href
  {\doibase 10.1109/TPS.2008.927143} {\bibfield  {journal} {\bibinfo  {journal}
  {IEEE Trans. Plasma Science}\ }\textbf {\bibinfo {volume} {36}},\ \bibinfo
  {pages} {1790} (\bibinfo {year} {2008})}\BibitemShut {NoStop}%
\bibitem [{\citenamefont {Sgattoni}\ \emph {et~al.}(2012)\citenamefont
  {Sgattoni}, \citenamefont {Londrillo}, \citenamefont {Macchi},\ and\
  \citenamefont {Passoni}}]{sgattoniPRE12}%
  \BibitemOpen
  \bibfield  {author} {\bibinfo {author} {\bibfnamefont {A.}~\bibnamefont
  {Sgattoni}}, \bibinfo {author} {\bibfnamefont {P.}~\bibnamefont {Londrillo}},
  \bibinfo {author} {\bibfnamefont {A.}~\bibnamefont {Macchi}}, \ and\ \bibinfo
  {author} {\bibfnamefont {M.}~\bibnamefont {Passoni}},\ }\href {\doibase
  10.1103/PhysRevE.85.036405} {\bibfield  {journal} {\bibinfo  {journal} {Phys.
  Rev. E}\ }\textbf {\bibinfo {volume} {85}},\ \bibinfo {pages} {036405}
  (\bibinfo {year} {2012})}\BibitemShut {NoStop}%
\bibitem [{\citenamefont {Bigongiari}(2012)}]{bigongiari-thesis}%
  \BibitemOpen
  \bibfield  {author} {\bibinfo {author} {\bibfnamefont {A.}~\bibnamefont
  {Bigongiari}},\ }\emph {\bibinfo {title} {High intensity laser-plasma grating
  interaction: surface wave excitation and particle acceleration}},\ \href
  {http://pastel.archives-ouvertes.fr/docs/00/75/83/55/PDF/alebigo_TESI_rappor%
teurs.pdf} {Ph.D. thesis},\ \bibinfo  {school} {Ecole Polytechnique} (\bibinfo
  {year} {2012})\BibitemShut {NoStop}%
\bibitem [{\citenamefont {Gaillard}\ \emph {et~al.}(2011)\citenamefont
  {Gaillard} \emph {et~al.}}]{gaillardPoP11}%
  \BibitemOpen
  \bibfield  {author} {\bibinfo {author} {\bibfnamefont {S.~A.}\ \bibnamefont
  {Gaillard}} \emph {et~al.},\ }\href {\doibase 10.1063/1.3575624} {\bibfield
  {journal} {\bibinfo  {journal} {Phys. Plasmas}\ }\textbf {\bibinfo {volume}
  {18}},\ \bibinfo {eid} {056710} (\bibinfo {year} {2011})}\BibitemShut
  {NoStop}%
\end{thebibliography}
\end{document}